%% file: main.tex
\documentclass[journal]{vgtc}                     

\onlineid{1060}

\vgtccategory{Research}

\title{From Instruction to Insight: Exploring the Functional and \\Semantic Roles of Text in Interactive Dashboards}

\author{%
  \authororcid{Nicole Sultanum}{0000-0001-8608-1427} and 
  \authororcid{Vidya Setlur}{0000-0003-3722-406X}
}

\authorfooter{
  \item
  	Nicole Sultanum and Vidya Setlur are with Tableau Research.
  	E-mail: \{nsultanum,vsetlur\}@tableau.com
}

\input{sections/0-abstract}


\teaser{
  \centering
  \includegraphics[width=.47\linewidth, alt={An analytical dashboard featuring a map and instructional text on the left side, and various bar charts on the right side.}]{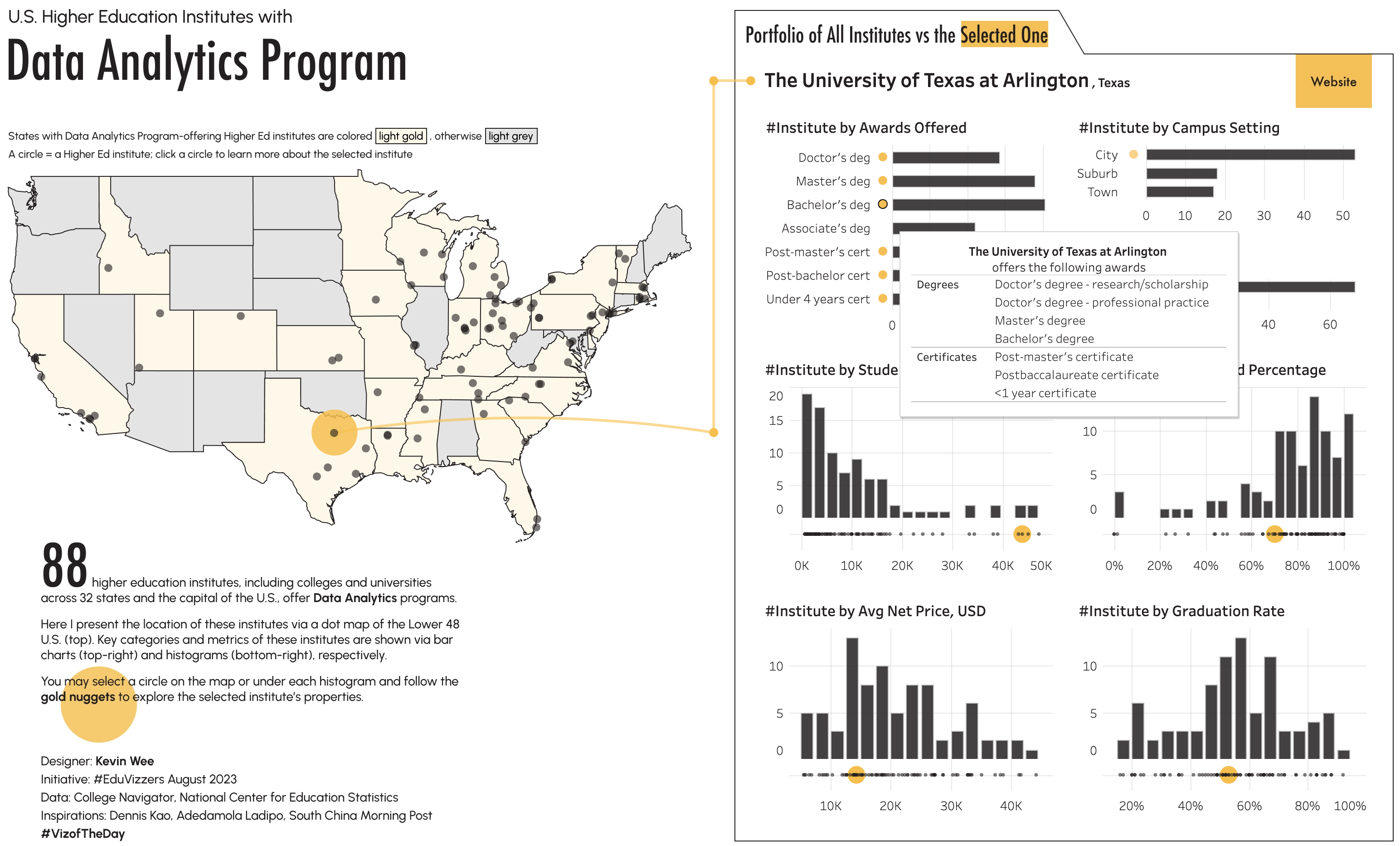}
  \hspace{1mm}
  \includegraphics[width=.51\linewidth, alt={An infographic style dashboard with a side panel, and a mix of text, charts, and descriptive statistics.}]{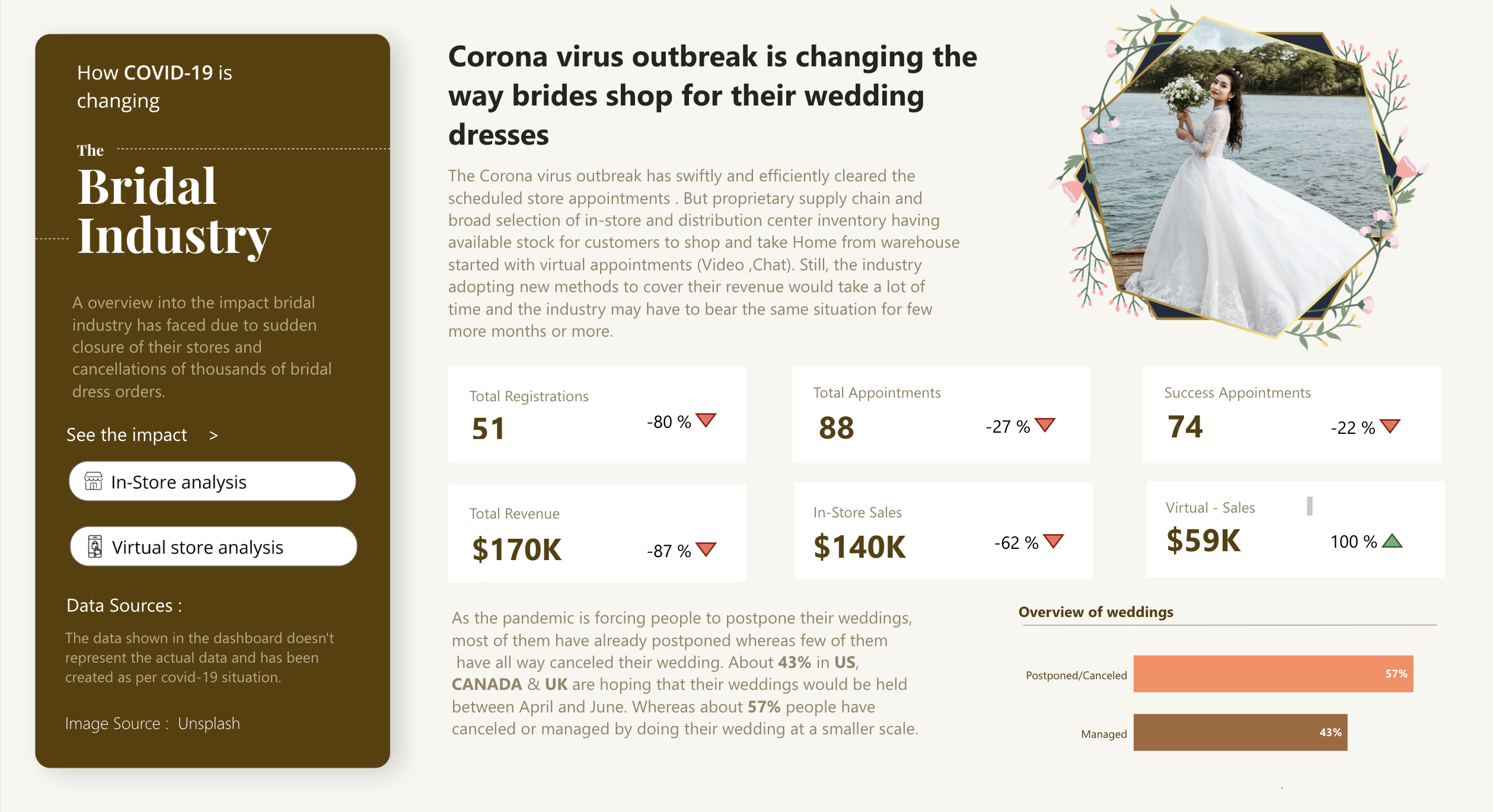}
  \caption{Examples of interactive dashboards that demonstrate how descriptive text elements can guide users through interactive data exploration and communicate key insights. Left \dbla{105}{https://public.tableau.com/app/profile/kevin.wee/viz/USHigherEdInstituteswithDataAnalyticsProgram/Dashboard}{Kevin Wee}{https://kevinwee.com/}: An analytic-style dashboard that provides descriptive and instructional text for the geographical distribution and profiles of U.S. higher education institutions with data analytics programs. Users can interact with the map to discover information about each institute, with supplementary bar charts and histograms providing details on the degrees offered, campus settings, student enrollment sizes, average net prices, and graduation rates. Right \dbla{173}{https://community.fabric.microsoft.com/t5/Data-Stories-Gallery/How-Covid-19-strikes-the-bridal-industry/td-p/1332158}{Infocepts}{https://www.infocepts.ai/}: An infographic-style dashboard showing changes in appointments, total revenue, and sales, with qualitative assessments to narrate the broader effects on shopping behavior and wedding planning. Visual elements, along with quantitative metrics and descriptive narratives, are employed to draw attention to data trends and provide an overview of the industry's adaptation to virtual platforms.}
  \label{fig:teaser}
}

\graphicspath{{figs/}{figures/}{pictures/}{images/}{./}} 

\usepackage{mathptmx}                  

\input{custom-imports-and-commands}

\begin{document}


\firstsection{Introduction}

\maketitle

\input{sections/01-intro}
\input{sections/02-relatedwork}

\input{sections/03-methodology}

\input{sections/04-typology}

\input{sections/05-heuristics}

\input{sections/06-future}
\input{sections/07-limitations}
\input{sections/08-conclusion}

\clearpage
\bibliographystyle{abbrv-doi-hyperref}

\bibliography{references}

\end{document}

%% file: sections/0-abstract.tex
\abstract{
There is increased interest in understanding the interplay between text and visuals in the field of data visualization. However, this attention has predominantly been on the use of text in standalone visualizations (such as text annotation overlays) or augmenting text stories supported by a series of independent views. In this paper, we shift from the traditional focus on single-chart annotations to characterize the nuanced but crucial communication role of text in the complex environment of interactive dashboards.
Through a survey and analysis of 190 dashboards in the wild, plus 13 expert interview sessions with experienced dashboard authors, we highlight the distinctive nature of text as an integral component of the dashboard experience, while delving into the categories, semantic levels, and functional roles of text, and exploring how these text elements are coalesced by dashboard authors to guide and inform dashboard users.
Our contributions are threefold. First, we distill qualitative and quantitative findings from our studies to characterize \textit{current practices} of text use in dashboards, including a categorization of text-based components and design patterns. 
Second, we leverage current practices and existing literature to propose, discuss, and validate \textit{recommended practices} for text in dashboards, embodied as a set of 12 heuristics that underscore the semantic and functional role of text in offering navigational cues, contextualizing data insights, supporting reading order, among other concerns.
Third, we reflect on our findings to identify gaps and propose \textit{opportunities} for data visualization researchers to push the boundaries on text usage for dashboards, from authoring support and interactivity to text generation and content personalization. Our research underscores the significance of elevating text as a first-class citizen in data visualization, and the need to support the inclusion of textual components and their interactive affordances in dashboard design.

}

%% file: custom-imports-and-commands.tex
\usepackage{multicol}
\usepackage{comment}
\usepackage{multirow}
\usepackage{enumitem}
\usepackage{makecell}
\usepackage{mdframed} 
\usepackage{enumitem}
\usepackage{soul}

\definecolor{newred}{RGB} {230, 70, 40}

\newcommand{\rev}[1]{#1}

\newcommand{\pheading}[1]{\vspace{4px}\noindent\textbf{#1}}

\definecolor{vlg}{RGB} {230, 235, 242} 
\newcommand{\typ}[1]{{\sethlcolor{vlg}\hl{\,\textmd{#1}\,}}}

\newcommand\itema{\item[$\rightarrow$]}
\newcommand\itemb{\item[$\Rightarrow$]}

\newcommand{\heurt}[1]{{\small {#1}}} 
\newcommand{\heid}[1]{\texttt{{#1}}} 

\definecolor{dbidc}{RGB} {13, 143, 119} 
\newcommand{\db}[1]{\textit{\textcolor{purple}{(\##1)}}}
\newcommand{\dbl}[2]{\textit{\href{#2}{\textcolor{purple}{(\##1)}}}}

\newcommand{\dbla}[4]{(by \textit{\href{#4}{#3}}, 
\textit{\href{#2}{\textcolor{purple}{\##1}}})}

\definecolor{vlg2}{RGB} {216,223,235}

\newcommand{\implications}[1]{
\begin{mdframed}[backgroundcolor=vlg, roundcorner=10pt, innertopmargin=3pt, innerbottommargin=3pt, skipabove=10pt, skipbelow=10pt,innerleftmargin=3pt,
linecolor= gray!60
]
\textbf{{Implications:}} {#1}
\end{mdframed}
}


%% file: sections/01-intro.tex
Dashboards have become a prevalent analytical artifact for visualizing and interacting with data across a variety of domains. From their original conception as glanceable visual monitoring systems for time-critical decision making~\cite{few:dashboarddesign,setlur2023heuristics}, dashboards have evolved into versatile \textit{data communication tools} that support both analytical and storytelling purposes~\cite{sarikaya2018we, bach2022dashboard}. This shift in utility has led to an extensive body of research aimed at cataloging and categorizing dashboards, focusing on their communication goals~\cite{sarikaya2018we}, dashboard intent~\cite{pandey2023medley}, their interaction mechanisms~\cite{tory2021finding}, and conversational design heuristics~\cite{setlur2023heuristics}, all aspects that underscore the growing importance of dashboards to support effective communication of data to their intended audience.

Among the diverse components that make up a dashboard, text elements such as titles, labels, captions, and data descriptions play a pivotal role in this communication~\cite{few:dashboarddesign,tufte2001visual,kosara2013storytelling,segel2010narrative}. 
An assessment of public dashboards in the wild found a majority (55\%) contained blocks of text~\cite{srinivasan2023toward}, stressing the prevalence of textual elements in dashboard design. 
There has been increased interest in studying relevant uses of text in the context of individual charts~\cite{Stokes2022StrikingAB,kim2021,kong:2014}, but text use in dashboards has received relatively little attention so far. 
And arguably, it is a timely concern. The current repertoire of dashboard generation tools focuses on automating chart and layout creation\rev{, placing} less emphasis on text generation and placement~\cite{bongshin:2015,hearst2023}. With the rise of large language models, this gap can be more easily filled, but without a solid understanding of practices and expectations for text, we are ill-equipped to design appropriate guardrails that can steer dashboard content generation in useful, meaningful directions.

Acknowledging the growing relevance of text in data visualization and the current landscape of generative AI for text, our work aims to understand and formalize the use of text in dashboards, and to provide guidance for dashboard authors and dashboard tool designers alike.
Our approach was informed by (a) a survey and qualitative analysis of 190 publicly available dashboards to catalog features of text and characterize their impact on dashboard design and communication, and (b) design review sessions with 13 dashboard creators to reflect and learn from on their personal practices for text use (\S\ref{sec:coding-methodology}). 
Our contributions are threefold. First, we (i) outline \textit{current practices} for text in dashboards based on dashboard analysis and expert feedback (\S\ref{sec:typology}), which helped inform a typology of text elements in dashboards and a compilation of text use patterns. 
Second, we discuss (ii) \textit{recommended practices} for text use in dashboards, based on an adaptation of Setlur et al.'s dashboard heuristics~\cite{setlur2023heuristics} and validated by experts (\S\ref{sec:heuristics}). 
Third, we outline (iii) \textit{future opportunities} for the usage of text in dashboards, reflecting on emerging challenges for text use and potential solutions based on the state of the art in data visualization research (\S\ref{sec:opportunities}). 
Our work sheds light onto the functional, structural, and semantic aspects \rev{of text} and reflects on its vital role in dashboards for analytical and navigation support. Our investigation shows that text is much more than a mere accompaniment to visuals but rather a fundamental enabler for narrative, explanation, and insight discovery.

%% file: sections/02-relatedwork.tex
\section{Related Work}
Our work builds on several lines of work: dashboard characteristics \& guidelines, text \& chart integration,
and dashboard authoring \& search.

\subsection{Dashboard Characteristics and Guidelines}
The role of dashboards has evolved to support the increasing complexity of user needs and the recognition that dashboards serve not just as a means for data consumption but as platforms for data conversation. Tory et al.'s~\cite{tory2021finding} concept of \textit{data conversations} highlights the diverse communication goals of dashboard users, emphasizing tasks such as summarizing, monitoring, and predicting insights. 

Dhanoa et al.\cite{dhanoa2022process} and Sarikaya et al.\cite{sarikaya2018we} identify key elements that contribute to an effective dashboard, including the importance of guiding users through the data exploration process with clear narratives and structured interaction pathways. Bach et al.~\cite{bach2022dashboard} extend this understanding by identifying design patterns and dashboard genres that cater to varied user intents and interaction styles, where text plays a key role in data narration and guiding user exploration. Setlur et al.~\cite{setlur2023heuristics} present a set of $39$ dashboard design heuristics grounded in the Gricean maxims of cooperative conversation, a framework that views dashboard interaction as a form of \textit{analytical conversation}. The heuristics consider the role of text in facilitating analytical conversation, ranging from legibility and descriptiveness, effective use of iconography, and the provision of navigational cues and breadcrumbs. Aspects of communication that dashboard text contributes to are acknowledged and studied in these works, but not how \rev{text} specifically influences this communication. 

Our work is also informed by a long history of design heuristics for interfaces and visualizations that include Nielsen's ten usability guidelines~\cite{nielsen1994enhancing, nielsen2005ten} and Shneiderman et al.'s eight golden rules for interface design and evaluation~\cite{shneiderman2016designing}, which provide \rev{directives} for user-centered design. The examination of usability heuristics for evaluating visualizations by Tory \& M\"{o}ller~\cite{tory2005evaluating} emphasizes \rev{the benefit} of incorporating both usability and visualization expertise in creating effective analytical interfaces. Subsequent work has aimed to develop and evaluate visualization-specific heuristics~\cite{forsell2010heuristic, vaataja2016information, tarrell2014toward, zuk2006heuristics, dowding2018development, cuttone2014four}, expanding upon the principles of user-centered design to cater specifically to the unique challenges and opportunities presented by visual data representation. 
Resources by Few~\cite{few:dashboarddesign}, Yigitbasioglu\cite{yigitbasioglu2012review}, and Wexler~\cite{ wexler2017big} introduce a user-centered design perspective to dashboard design, complemented by frameworks that focus on user goals and intents~\cite{lam2017bridging,lee2022affective}. Together, these works guide visualization design towards more meaningful user experiences and support users' cognitive processes~\cite{warebook:2008}.

\rev{We build} upon this body of research \rev{by examining} the role of text in dashboards along with its categories, semantic levels, and functional roles in guiding and informing users. Through an extensive \rev{dashboard coding exercise}, we uncover \rev{the interplay} between text and dashboard visualizations to \rev{support analytical} and narrative communication goals. 

\subsection{Text and Chart Integration}
Recent research has explored the integration of text and charts in visualizations through various lenses, offering insights into how text can influence the perception and understanding of visualized data.

Kim et al.~\cite{kim2021} conducted a crowdsourced study to understand how readers synthesize information from both charts and captions, finding that the emphasis on high-prominence features within both elements led readers to identify those features as key takeaways. Their research underscores the importance of coherence between visual and textual elements and how external context can enhance the reader's comprehension of the chart's message. Building on these insights, Lundgard and Satyanarayan~\cite{lundgard2021accessible} proposed a four-level semantic categorization of text content designed to enhance visualization accessibility. Their framework distinguishes between perceiver-independent descriptions, such as objective chart specifications, and perceiver-dependent insights, including interpreted observations and contextual knowledge. Recent work broadens the scope of accessible and inclusive data representation by championing the parity of visualization, textual description, and sonification in multimodal data analysis~\cite{zong2024umwelt}.

Further exploring the role of text in visualizations, Stokes~\cite{Stokes2022StrikingAB} observed that study participants favored heavily annotated charts over less annotated charts or text alone. This preference highlights the added value of text in aiding data interpretation, with emphasis on how different types of semantic content\,—\,statistical, relational, elemental, or encoded\,—\,impact the takeaways drawn by readers. This finding aligns with Stokes and Hearst's advocacy for treating text as co-equal to visualization, urging researchers to focus on the readability and integration between these two modalities~\cite{stokesgive}. Their call to action reflects a growing body of work, including Ottley et al.~\cite{ottley2019curious}, exploring the critical role of text in visual analysis and its importance for conveying key messages to the reader. Brath provides for a more integrated and text-centric perspective on visualization, suggesting new avenues for enhancing data communication through textual integration~\cite{brath2020visualizing}. 

Eye-tracking studies by Borkin et al.~\cite{borkin2015beyond} have shown that participants were more likely to fixate on and recall textual content surrounding visualizations, such as titles and labels. Similarly, Kong et al.~\cite{kong2018frames,kong2019trust} investigated the influence of titles on the perceived message of visualizations, discovering that slanted framings (e.g., emphasizing only part of the chart's message) significantly impacted recall and interpretation. \rev{Zhi et al.~\cite{zhi2019linking} found that explicit text-chart integration in data-driven stories may increase recall and engagement.
Several tools build on these synergies, e.g., FacetNotes~\cite{badam2022integrating}, a novel concept for integrating text annotations directly with data points on interactive dashboards; Kori~\cite{latif2021kori}, a tool to support identification of text-chart links; and VizFlow~\cite{sultanum2021leveraging}, a tool that leverages text-chart links to generate dynamic data-driven articles.} 

However, the preference for integrating text and visual elements is not always universal. Hearst \& Tory~\cite{hearst2019would} investigated visualization preferences in conversational interfaces with chatbots and found a notable division in participant preferences. Nearly half of the participants expressed a preference against viewing charts, while those who did prefer charts also favored the inclusion of additional contextual data within the visual representation.  
Based on these foundational studies, we specifically focus on the dynamic interplay between text and chart features in the context of interactive dashboards. Unlike previous research that primarily considers static visualizations or singular aspects of text integration, our work investigates how authors specifically incorporate text elements in dashboards, exploring both the role of text and their various forms. 




\subsection{Dashboard Authoring and Search}
\rev{Understanding the analytical intent for authoring dashboards is crucial for effectively integrating text and chats for enhancing the overall communicative effectiveness of dashboards~\cite{brehmer:2013,schulz:2013}.} Recent work by Pandey et al.~\cite{pandey2023medley} proposed a set of dashboard intents and objectives accompanied by specific objectives like summarizing measures, comparing categories across dimensions, and displaying univariate summaries, to support dashboard composition through auto-generated variations of intent and objectives. 
Srinivasan and Setlur~\cite{srinivasan2023bolt} extend the capability for dashboard composition through a natural language interface, enabling authors to generate and arrange dashboard views based on user-provided linguistic utterances. 
Machine learning techniques have also been explored for dashboard authoring and search. Ma et al.\cite{ma2020ladv} and Wu et al.\cite{wu2021multivision} propose deep learning-based methods to assist in dashboard generation and chart view suggestions, respectively, demonstrating the potential of ML to expedite dashboard prototyping and enhance the generation process. Similarly, Deng et al.~\cite{deng2022dashbot} introduce DashBot, leveraging reinforcement learning to automate dashboard generation, focusing on the creation of valid dashboards that facilitate data insight discovery and support direct manipulation of recommendations. However, this prior work tends to focus less on the systematic integration of text beyond basic labels and titles in dashboard generation, presenting an opportunity to explore the role of text to further enhance the dashboard's communicative effectiveness. 

Oppermann et al.~\cite{oppermann2020vizcommender} explore leveraging text-based content within dashboards to aid in dashboard search and recommendation for specific analytical questions. Their work highlights the challenge of sparse text presence in dashboards and the necessity of including non-visible text for context, thus offering a pathway toward more meaningful recommendations based on textual content. 
EmphasisChecker~\cite{Kim2023} explores the integration of text and charts more directly in the authoring experience by identifying mismatches between the visual emphasis of charts and the textual emphasis of captions. 
Our work extends beyond prior research that primarily focused on dashboard intents, objectives, and composition; rather, we explore the specific interplay between text and visualizations, wherein text is not merely an accessory to visual elements but a primary medium for communication within dashboards, conveying key insights to the intended audience.

%% file: sections/03-methodology.tex
\section{Methodology}
\label{sec:methodology}

Our efforts to characterize text use in dashboards are threefold: to identify relevant features of current text use, to provide recommendations for text use, and to identify opportunities for improvement. To support these goals, we performed two elicitation studies. First, we surveyed and analyzed a large collection of publicly available dashboards, which informed a typology of dashboard text elements and provided samples to illustrate best practices. Second, we conducted interviews with 13 dashboard designers, which helped contextualize and validate findings from our dashboard survey and inform challenges of text use in dashboards, which in turn helped inform research opportunities for text use in dashboards. We detail our study methodology in the following subsections, and our cross-cut findings are presented throughout our investigation goals to describe \textit{current practices} (\S\ref{sec:typology}), \textit{recommended practices} (\S\ref{sec:heuristics}), and \textit{opportunities} (\S\ref{sec:opportunities}).

\subsection{Dashboard Corpus}
\label{sec:dashboard-corpus}

We compiled a large and diverse collection of 190 publicly available dashboards (Tableau~(116), Power BI~(50), and other web visualization tools~(24)) sourced four ways: 

\pheading{1. Dashboard corpora from prior work} (52).
We included dashboards listed in prior surveys~\cite{bach2022dashboard, sarikaya2018we, setlur2023heuristics} that were both in English and publicly available, yielding 52 dashboards (from a total of 173). While these dashboards cover a relatively diverse range of platforms and genres, they have been opportunistically sampled, which could suggest a skewed selection of dashboards from limited sources and topics. 

\pheading{2. Curated sample from Tableau Public}~(74).
To broaden diversity, we gathered dashboards from a large dataset of \rev{Tableau Public~\cite{tableaupublic} dashboards} featured in Srinivasan et al.~\cite{srinivasan2023toward} (with assistance from the authors). 
From an initial random sample of dashboards, we filtered for dashboards in English featuring five or more dashboard zones (i.e., charts, filters, legends, text, multimedia blocks), with at least one text zone and one chart zone. 
This led to a diverse set of 74 dashboards, spanning a 
broad range of genres, topics, communication goals, analytical complexity, and levels of refinement. 

\pheading{3. Popular Power BI dashboards} (43).
We also included a third set of 43 Power BI dashboards, particularly those highly regarded in the ``Top Kudos'' category (i.e., high community value) of the \rev{Power BI Data Stories Galleries~\cite{powerbigallery}} to ensure platform diversity and to analyze best practices in dashboard design across different software ecosystems. 

\pheading{4. Dashboards from expert interview participants }(21).
We included a fourth set of 21 dashboards from interviews (\S\ref{sec:interview-methodology}) that were authored by expert interview participants and handpicked by them as meaningful examples of text use in their dashboard practice.  While most (19) are Tableau dashboards, they present a balanced coverage of genres and topics, from business dashboards to infographics for the general public.

\subsection{Dashboard Analysis}
\label{sec:coding-methodology}

The first author conducted an initial round of inductive analysis on a subset of dashboards to flesh out a preliminary typology of text-based components. \rev{Both} authors were then involved in a follow-up deductive stage to refine these codes and incorporate other text-relevant features from past literature that encode dashboard semantics, including \textit{dashboard genres}~\cite{bach2022dashboard}, 
\textit{dashboard goals}~\cite{sarikaya2018we} 
and \textit{semantic levels of data-driven content}~\cite{lundgard2021accessible}. This resulted in a code book for independent coding covering a typology of text components and relevant dashboard-level~features.  
A detailed breakdown of our derived typology is presented in \S\ref{sec:typology}. 

We also compiled a set of heuristics for text usage in dashboards. Based on the work by Setlur et al.~\cite{setlur2023heuristics} that proposes 39 heuristics for dashboard design to foster analytical conversation,
we reframed their heuristics from a perspective of text use, leading to a set of 12 heuristics for text use.
These heuristics were appended to the code book alongside a detailed rubric listing 4 levels of application for each heuristic: \textit{strong application}, \textit{weak application}, \textit{weak violation}, \textit{strong violation}. A detailed discussion on the derived heuristics is presented~in~\S\ref{sec:heuristics}.

Using the code book, \rev{both} authors conducted an independent assessment of dashboards in our corpus to extract meaningful text features and assess the application of the heuristics. Authors met regularly during the coding process to resolve disagreements and refine code book descriptions and rubric. In parallel, recurrent and interesting applications of text use were also cataloged. Unanimous consensus was reached on all features for the 190 dashboards in our collection. 

Outcomes of this coding effort showcased the diversity of our corpus, featuring a broad range of dashboard \textit{genres}~\cite{bach2022dashboard} (\textit{analytic}~(59\%), \textit{infographic}~(25\%), \textit{static}~(10\%), \textit{repository}~(8\%) and \textit{magazine}~(4\%)) and dashboard \textit{goals}~\cite{sarikaya2018we} (\textit{decision making}~(40\%), \textit{awareness}~(22\%), \textit{motivation \& learning}~(38\%)). 
It also contributed a more systematic look over patterns of text use (\S\ref{sec:typology}), meaningful examples to showcase patterns (\S\ref{sec:heuristics}), and inspiration for future opportunities for text use in dashboards (\S\ref{sec:opportunities}). 
Our coded dashboard corpus and code book can be found in Supplemental Materials. We reference specific dashboard instances by their corpus \db{ID} (with embedded access links). 

\input{sections/04-fig-typology}
\subsection{Expert Interviews}
\label{sec:interview-methodology}

To complement our dashboard analysis efforts, we conducted interview sessions with experienced dashboard creators. Our goal was to validate observations of text use from the dashboard analysis, elicit rationale for design patterns, document common practices for text use, and validate our proposed set of heuristics. 

We recruited 13 participants ($P1$-$P13$) from various channels, including social media, specialized community forums, targeted recruitment, and word of mouth. Most participants create dashboards professionally; listed occupations include business intelligence (BI) managers, data analysts, BI consultants, and students in data-relevant fields. All participants had at least one year of experience producing dashboards, with 10 reporting at least five years of professional experience on the topic. Tableau (11/13) and Power BI (5/13) were the most commonly cited dashboard platforms, but several participants also used other tools (5/13), and many stated being conversant in more than one tool (6/13). 

Sessions were one hour long and consisted of a semi-structured interview on text practices for dashboards followed by a feedback segment on the heuristics. Prior to the session, participants were asked to provide links to dashboards they authored that made for interesting discussion on the use of text (and were later included in our fourth sample set, \S\ref{sec:dashboard-corpus}). At the session, we asked them to share design rationale on text use for \rev{their dashboards}, including semantics, level of detail, text formatting choices, layout and reading order, user guidance, dynamic text, and \rev{accessibility}. We also inquired about practices and challenges they \rev{generally face with text}. We then presented our list of heuristics and asked them to reflect on how representative they were of their current practices and what aspects of \rev{text} practices we should further consider. Sessions took place over video conference and were screen-recorded and auto-transcribed. Findings from transcripts were derived via thematic analysis and contributed to contextualize patterns of dashboard text use (\S\ref{sec:typology}), validate our proposed heuristics~(\S\ref{sec:heuristics}), and inform opportunities for text use in dashboards (\S\ref{sec:opportunities}).

%% file: sections/04-fig-typology.tex
\begin{figure*}[ht]
\centering
 \begin{subfigure}[b]{0.45\textwidth}
 \includegraphics[width=\textwidth]{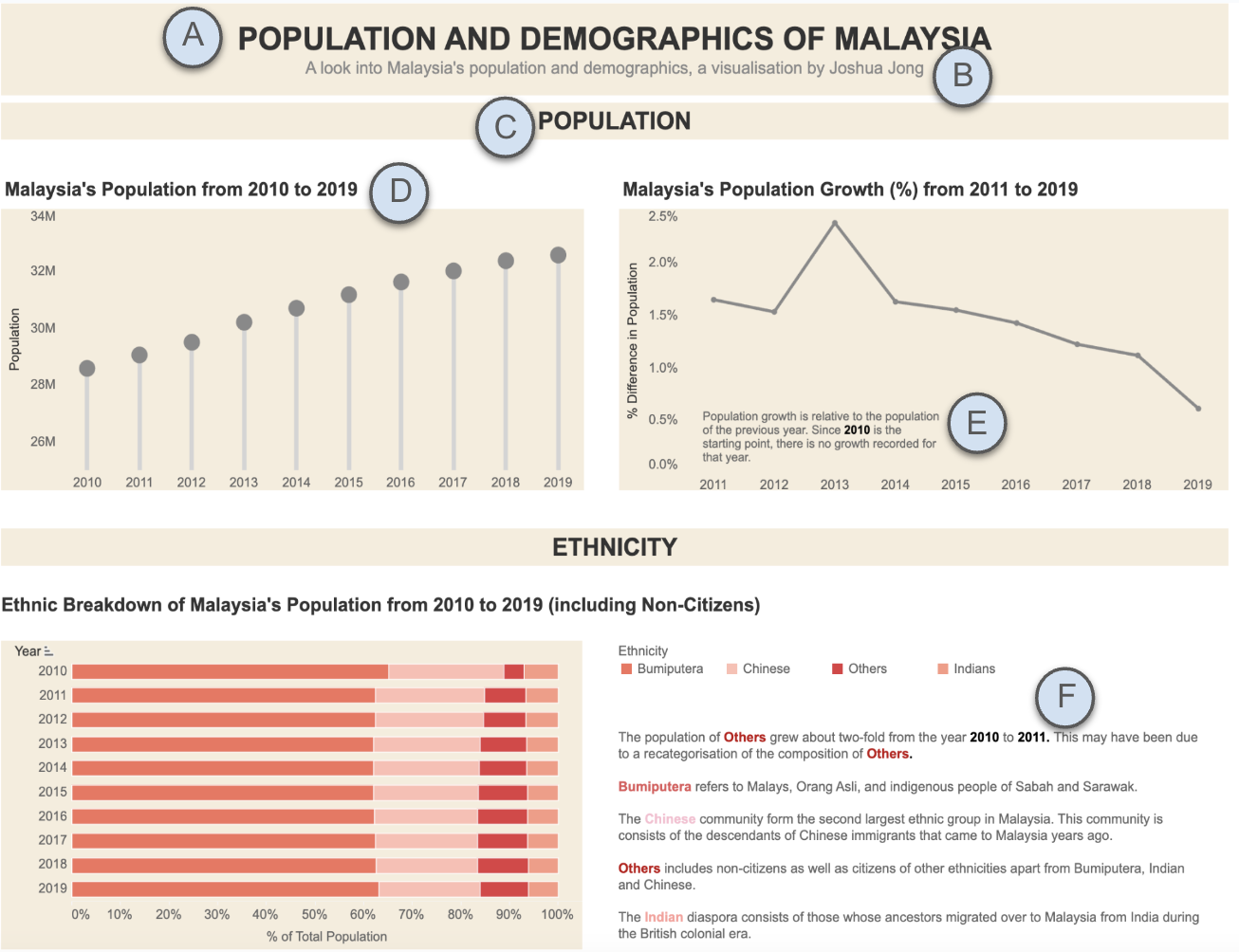}
  \caption{A dashboard visualizing the population and demographic trends in Malaysia.}
\label{fig:malaysia_dashboard}
\end{subfigure}
 \begin{subfigure}[b]{0.5\textwidth}
\includegraphics[width=\textwidth]{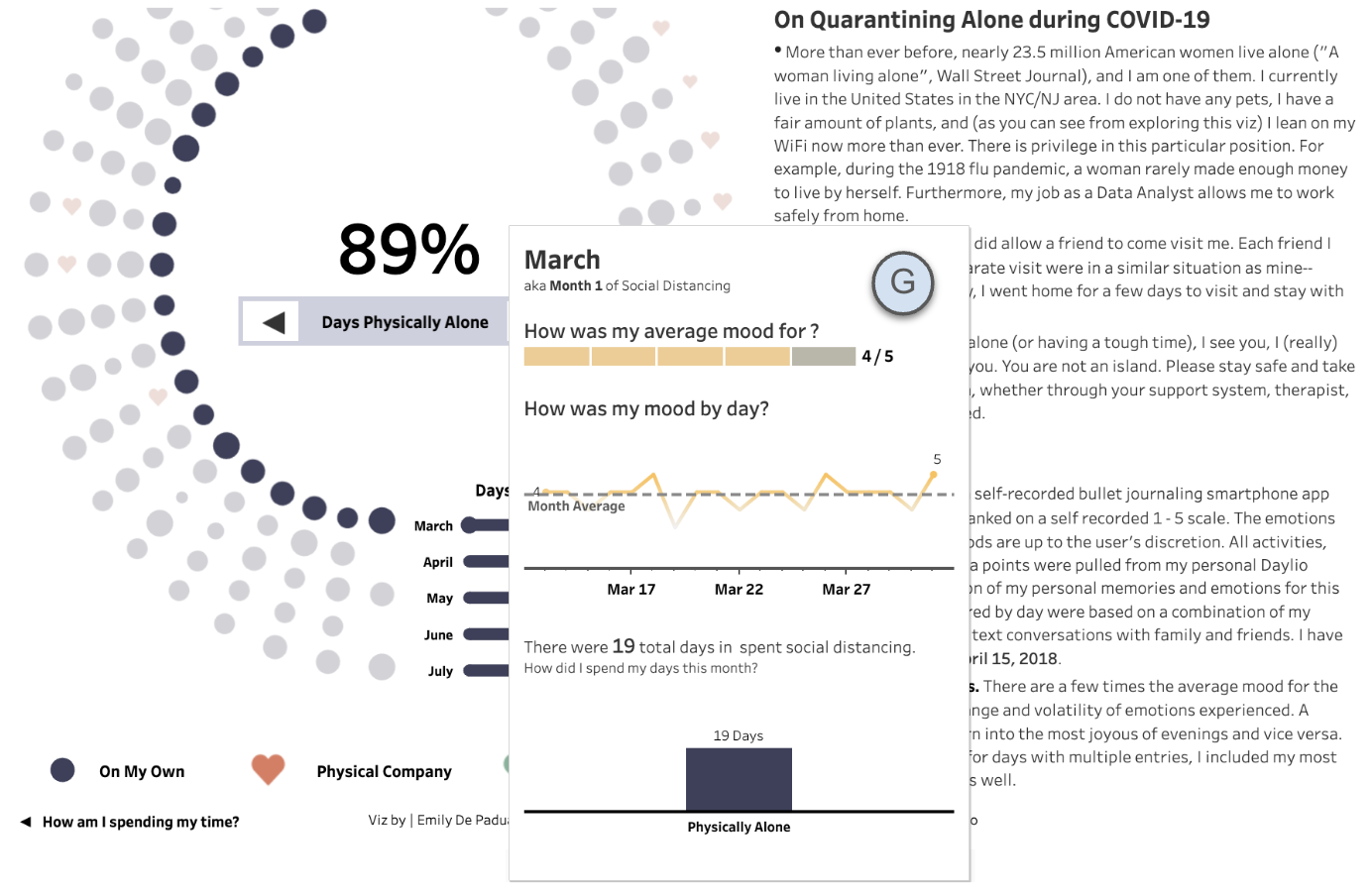}
\caption{A personal dashboard tracking the experience of social distancing during the COVID-19 pandemic.}
\label{fig:social_distancing_dashboard}
\end{subfigure}
\vspace{1em} 
 \begin{subfigure}[b]{0.45\textwidth}
\includegraphics[width=\textwidth]{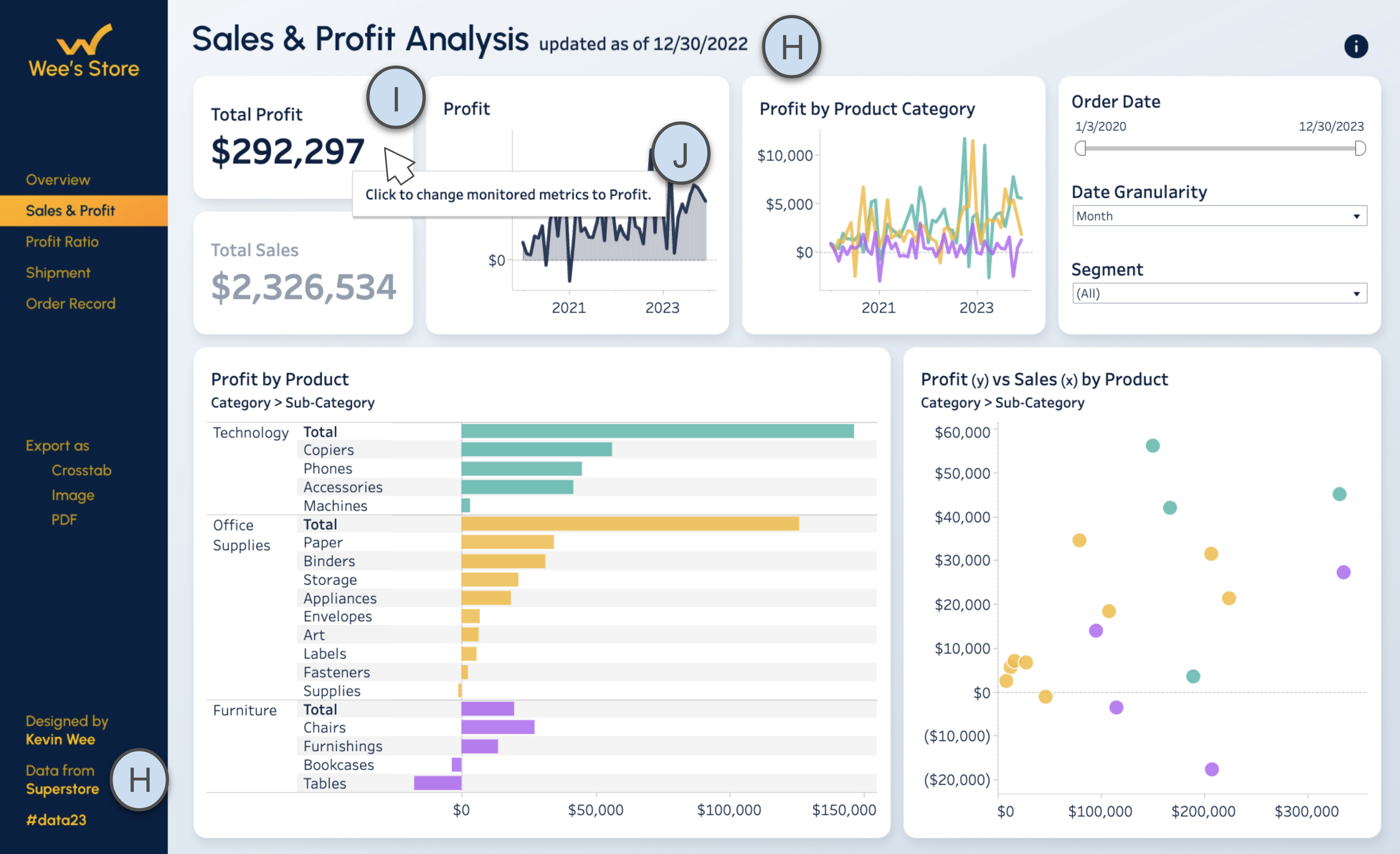}
\caption{A business dashboard for sales and profit analysis.}
\label{fig:transport_dashboard}
\end{subfigure}
\begin{subfigure}[b]{0.5\textwidth}
\includegraphics[width=\textwidth]{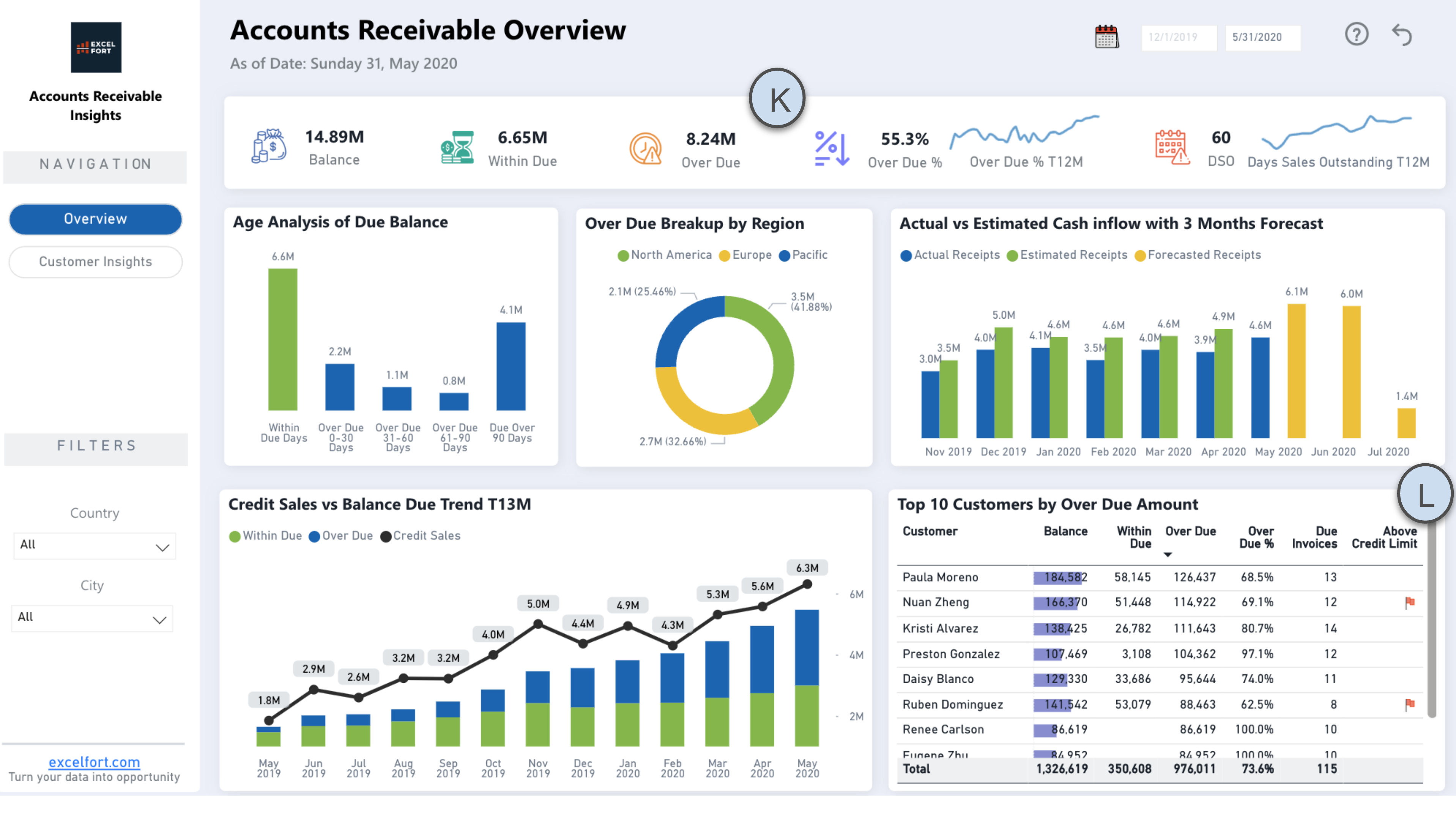}
\caption{A business dashboard for sales monitoring.}
\label{fig:school_finance_dashboard}
\end{subfigure}
\caption{Text components in various dashboard snippets. 
Figure 2(a) \dbla{87}{https://public.tableau.com/views/assignment1_16295362252850/Dashboard13}{Joshua Jong}{https://public.tableau.com/app/profile/joshua.jong6344/vizzes}: (A) The title provides an overarching theme; (B) The subheading offers additional context and credits the creator; (C) Section headers categorize the dashboard into thematic areas; (D) chart titles denote the focus of each visualization; (E) chart annotations explain specific data points or trends for clarity; and (F) Content blocks supply narrative insights, interpreting the ethnic composition changes over time and providing definitions for terms such as `Bumiputera' and the 'Indian' diaspora. 
Figure 2(b) \dbla{136}{https://public.tableau.com/app/profile/emily.de.padua/viz/AllByMyself/AllbyMyself}{Emily De Padua}{https://www.emilydepadua.com}: Features a prominent percentage highlighting days spent alone, a narrative text block reflecting personal reflections on quarantine, and a (G) custom tooltip displayed on hover, showing an interactive timeline charting average daily mood. 
Figure 2(c)  \dbla{59}{https://public.tableau.com/app/profile/kevin.wee/viz/Data23DemoCreativeUsesofImagesinaTableauDashboard_16828225482900/Overview}{Kevin Wee}{https://kevinwee.com}: (H) Metadata blocks, indicating author, data source, and when the dashboard was last updated; (I) text-data summary blocks featuring total profit and sales, and (J) interaction guidance on tooltip hover.
Figure~2(d) \dbla{156}{https://community.fabric.microsoft.com/t5/Data-Stories-Gallery/Accounts-Receivable-Dashboard/td-p/2530321}{Fowmy Abdulmuttalib}{https://excelfort.com}:~(K) A strip of text-data summaries featuring icons; and (L) a data table featuring embedded graphics.
}
\label{fig:texttypes}
\vspace{-1em}
\end{figure*}

%% file: sections/04-typology.tex
\input{sections/04-fig-semantic-levels}
\section{Characterizing Current Text Practices}
\label{sec:typology}


Our examination of current practices for dashboard text is based on findings from our coding exercise (\S\ref{sec:coding-methodology}) and interviews with expert dashboard creators (\S\ref{sec:interview-methodology}). 
First, we provide a characterization of text components in dashboards from a \textit{functional} lens by outlining different embodiments and roles of text use (\S\ref{sec:text-components}) and from a \textit{semantic} lens via an adaptation of Lundgard and Satyanarayan's four-level model of semantic content~\cite{lundgard2021accessible} (\S\ref{sec:semantic-levels}).  We then discuss prominent \rev{text usage patterns} that emerged from the coding exercise and expert interviews, presented alongside implications for design (\S\ref{sec:design-patterns}).


\subsection{Text Components}
\label{sec:text-components}
We identified two groups of dashboard components where text plays an integral role: (a) \textit{text blocks}, i.e., where text appears as a standalone element; and (b) \textit{text-based components}, where text is intrinsically combined with visual and interactive features but \rev{ is still prominent}. Text decorators to interactive widgets (e.g., filter labels, drop-down labels) and primarily visual components (e.g., chart axis and ticks and color legend labels) are not considered in this assessment, 
\rev{as their content and functionality are} 
dictated by design requirements of \rev{corresponding} charts and legends, 
rather than offering independent or additional narrative or analytical insights. 
We also did not explicitly code for text visualization instances, as their use of text is highly bespoke.

The list of \textit{text blocks} we identified include: 

\begin{itemize}
    \itema \typ{Title} (present in 92\% of dashboards in our corpus). A prominent, discernible label (or short sentence) displayed with a font size larger than other text blocks in the dashboard (Fig. \ref{fig:texttypes}(A)). 
    \itema \typ{Subheading} (36\%). Text that goes below (or beside) the dashboard title and adds more context to the title (Fig. \ref{fig:texttypes}(B)). 
    \itema \typ{Section Header} (29\%). A textual element that labels a partition of a dashboard containing multiple components (e.g., a text block alongside one or more views) and provides a high-level overview of what the ensuing section contains (Fig. \ref{fig:texttypes}(C)).
    \itema \typ{Chart Title} (84\%). Text labeling an individual chart or view (Fig. \ref{fig:texttypes}(D)). Even though they play a supportive role to charts, they allow for more authoring freedom than text decorators and are thus included in our assessment as a standalone component.
    \itema \typ{Interaction Guidance} (58\%). Text that describes possible interactions with the dashboard, often written in imperative language, e.g., ``\textit{click here to filter}'' (Fig. \ref{fig:texttypes}(J)).
    \itema \typ{Metadata} (76\%). A feature borrowed from Bach et al.~\cite{bach2022dashboard}, it designates text that describes data sources (53\%), author (43\%), last update (18\%), and data-related disclaimers (26\%) (Fig. \ref{fig:texttypes}(H)).
    \itema \typ{Content Block} (67\%). Sentence or multiline text that (a) supports the analytical and narrative goals of the dashboard (such as explaining what the data is, the visual encodings, salient points, and relevant context), and (b) is hierarchically perceived as ``\textit{body}'' level text (Fig. \ref{fig:texttypes}(F)).
\end{itemize}

The list of \textit{text-based components} that we consider are: 
\begin{itemize}
    \itemb \typ{Text-data summaries} (58\%). Also known as BANs (i.e., ``\textit{big-a** numbers}''), they feature key measures (i.e., a numerical value and a corresponding label) in a prominent way (Fig. \ref{fig:texttypes}{(I)} and {(K)}). 
    \itemb \typ{Chart Annotation} {(19\%)}. Text callouts overlaid on charts to bring attention to particular data elements, e.g., salient points, trends, data-specific context, and interaction guidance (Fig. \ref{fig:texttypes}{(E)}).
    \itemb \typ{Data Table} {(35\%)}. A (still very prominent) text-based alternative to showcase data (Fig. \ref{fig:texttypes}{(L)}). 
    \itemb \typ{Tooltip}. A dynamic text callout that appears when hovering data marks (Fig. \ref{fig:texttypes}{(G) and {(J)}}). Dashboards created in Tableau and Power BI implement basic default tooltips that show the numerical values for the hovered data points, and so they are nearly ubiquitous (i.e., appearing in nearly 100\% of our dashboards). However, a substantial subset of them {(42\%)} are customized to display text content in a richer way (e.g., in the form of sentences), making tooltips a relevant content authoring resource.
\end{itemize}


\subsection{Semantic Levels}
\label{sec:semantic-levels}

Beyond a functional understanding of text components,  our coding also sought to capture how text content shapes dashboard communication. To this end, we adapted Lundgard and Satyanarayan's four-level model for semantic content~\cite{lundgard2021accessible}, which was originally created in the context of accessibility to categorize analytical levels of detail of chart caption paragraphs.  
Given that the scope of our coding encompasses shorter, label-style text as opposed to fully formed sentences, we made a few adaptations when applying the four-level model to dashboard text components,  described below and exemplified in Fig.~\ref{fig:semanticlevels}:
\begin{itemize}
    \item \textbf{Level 1 (LV1): \textit{Elemental and encoded properties}} {(97\%)}. Refers to the description of charts and visible chart data, e.g., adjacent \typ{content blocks}, and \typ{chart titles}.
    \item \textbf{Level 2 (LV2): \textit{Statistical concepts and relations}} {(74\%)}. Refers to mentions of statistical features of the data formulated at a ``data facts''~\cite{srinivasan:2018} level, e.g., descriptive statistics, outliers, point-wise comparisons. \typ{Text-data summaries} generally fit in this category as they often depict sums, averages, and extrema in a similar granularity as LV2 statements.
    \item \textbf{Level 3 (LV3): \textit{Perceptual and cognitive phenomena}} {(27\%)}. Refers to more complex data statements, such as higher-level trends and pattern synthesis, while using more ``natural-sounding'' language than LV2 statements. \rev{Most often found in \typ{content blocks}, and occasionally \typ{chart annotations} and \typ{subheadings}}.
    \item \textbf{Level 4 (LV4): \textit{Contextual and domain-specific insights}} {(50\%)}. Text that provides additional context to the observed patterns and trends, e.g., historical, social, and political factors. \rev{Largely present in \typ{content blocks}, and sometimes \typ{tooltips}}. 
\end{itemize}

\subsection{Outlining Dashboard Text Practices}
\label{sec:design-patterns}

Building on the functional and semantic categorizations above, plus dashboard-level features~\cite{bach2022dashboard, sarikaya2018we} and expert interviews, we discuss various usage patterns for dashboard text alongside implications for design.
 
\subsubsection{Dashboard Text Practices are Based on Experience}
\label{sec:practices-experience}

When asked about their text use more generally, most participants (8/13) stated that they are not explicitly formalized ($P1$, $P4$, $P6$, $P8$) nor taught ($P6$), but rather based on experience, exposure, and honed senses, e.g., learning from others ($P6$, $P7$, $P13$),  learning on the job ($P5$, $P12$),  and developing a feel for ``\textit{what looks good}'' to them ($P3$, $P7$), which feeds into an implicit set of shared practices: ``\textit{even if it hasn't been formalized, there is kind of this collective understanding (..) because you've seen it repeated in other forms of conception}'' ($P4$). The more explicit guidance emerging from expert sessions was on structure and formatting guidelines borrowed from other established fields, such as publishing \& typography (e.g., magazines) ($P1$, $P4$, $P7$),  and graphic \& web design ($P5$, $P7$, $P13$). 

\implications{While text practices are not formalized, shared practices do exist and borrow inspiration from other communication and UX fields. Explicitly formalizing text practices is valuable and viable to some extent, but the most sustainable path forward is likely learning from examples (e.g., machine-learning approaches).}

\subsubsection{Text Components and Semantic Levels}
\label{sec:practices-semantic}

Our coding exercise helped \rev{inform} how text components tend to cover semantic levels. \typ{Chart titles} were the most \rev{frequent} contributors of LV1 content, and \typ{text-data summaries} of LV2 content. \typ{Content blocks} encompassed a lot of the higher level, fine-grained detail and context in dashboards, appearing in {96\%} of dashboards featuring LV3 content, and {90\%} of dashboards featuring LV4 content (versus {67\%} of all dashboards);  and so did \typ{chart annotations} (present in {42\%} of dashboards with LV3 and {28\%} with LV4 content, versus {19\%} of all dashboards).

That said, text components also showcase a lot of diversity in terms of the range of semantic levels they may support.  \typ{Tooltip} instances in our corpus expanded far beyond their default role of displaying data values under hovered marks (LV1),  to also correlate marks with other data dimensions (Fig.~\ref{fig:social_distancing_dashboard} 
\dbl{136}{https://public.tableau.com/app/profile/emily.de.padua/viz/AllByMyself/AllbyMyself}), list associated data facts 
\dbl{137}{https://public.tableau.com/views/TheCocktailCornersoftheWorld_IronViz_2021/TheCocktailCornersoftheWorld_Data}, add in historical context 
\dbl{55}{https://public.tableau.com/app/profile/diego.parker/viz/AnodetoChileanwine/Anodetowine}, and showcase disclaimers~\dbl{34}{https://public.tableau.com/app/profile/sam.epley/viz/BestStatestoRetireintheUnitedStates/BestStatestoRetireintheUnitedStates}. \typ{Chart annotation} content was found to be equally diverse, covering the whole range of LV1 (e.g., overlaying associated data values, 
\dbl{126}{https://public.tableau.com/views/DIH5_28V3/cmsent}), LV2 (e.g., emphasizing salient points and trends, 
\dbl{111}{https://public.tableau.com/views/TablesTest2/Dashboard} 
\dbl{118}{https://public.tableau.com/views/HomeworkWeek4_16291731666530/Dashboard1}), LV3 (e.g., explaining complex trends, 
\dbl{87}{https://public.tableau.com/views/assignment1_16295362252850/Dashboard13} 
\dbl{37}{https://public.tableau.com/app/profile/fuadahmed/viz/TheCostofLiving/Dashboard1xc}) and LV4 content (e.g., incorporating external context, 
\dbl{79}{https://public.tableau.com/views/Hospitalizations_age_Dec2020_BD/Dashboard2-Age} 
\dbl{58}{https://public.tableau.com/views/SDGVizProject12/SDGS12}). We also noted several instances of \typ{titles} and \typ{subheadings} being used to convey LV3 content, such as summarizing higher-level insights from the data and framing them as takeaways, e.g., 
\dbl{178}{https://community.fabric.microsoft.com/t5/Data-Stories-Gallery/Analyzing-the-UN-s-Sustainable-Development-Goals-at-Gartner-Data/td-p/2527235},   Fig.~\ref{fig:teaser} (right) 
\dbl{173}{https://community.fabric.microsoft.com/t5/Data-Stories-Gallery/How-Covid-19-strikes-the-bridal-industry/td-p/1332158}. 

Related to this last observation, it is worth noting that the hierarchical structure of text blocks (\typ{titles}, \typ{subheading}, \typ{section headers}, \typ{chart titles}) also tend to mirror a hierarchy of semantic concepts, going from \typ{titles} and \typ{subheadings} that allude to higher level analytical goals down to chart titles which are closer to the data. 






\implications{There is a strong relationship between structural levels of text and their semantic level coverage. This underscores that the visual design and formatting of text is more than an aesthetic concern. From a content generation perspective, this means not only looking at semantic hierarchy, but also how that hierarchy may be scaffolded by an explicit, visual text structure.}








\begin{figure}
    \centering
    \includegraphics[width=0.9\columnwidth, alt={Level 2 content in a dashboard tooltip: "Perception (2) (Harvey Walters) is found in 52\% of Amanda Sharpe decks (51 out of 97)". Level 3 content in the same tooltip: "That is 0.04 decks per day, or 1 deck every 25.6 days, based on the most recent of either the card release date (Sep '20) or investigator release date (Sep '20)".}]{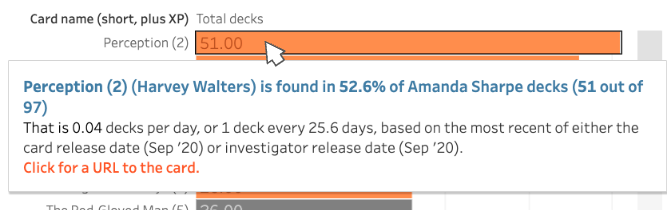}
    \caption{Tooltip featuring LV2-LV3 content  \dbla{77}{https://public.tableau.com/app/profile/acotgreave.tableausoftware.com/viz/ArkhamDBDataDrivenDeckBuidling/DeckBuilderpopularcards2}{Andy Cotgreave}{https://uk.linkedin.com/in/acotgreave}.}
    \label{fig:sentence-tooltip}
\vspace{-1em}
\end{figure}

\subsubsection{Text Use Across Dashboard Genres \& Goals}
\label{sec:practices-genres}

The strongest signal we found to discriminate text use, based on the coding exercise and on expert feedback, is if the dashboard is more narrative-driven (e.g., \textit{infographic} and \textit{magazine} genres, often \textit{motivation \& learning} goal, for {general audiences}, casual use), or more exploratory (e.g., largely \textit{analytic} genre, and \textit{decision making} goals, for {expert audiences}, recurrent use). 

By definition, narrative-driven dashboards make far more liberal use of explicit salient points and takeaways. This emphasis is clear when we look at the prevalence of perceiver-dependent content (LV3-LV4) and associated text components against \textit{infographic} (more explanatory) and \textit{analytic} (more exploratory) dashboards. Infographics feature more \typ{content blocks} ({91\%},  vs. {53\%}), \typ{chart annotations}  ({29\%} vs. {14\%}), and LV3-LV4  content ({57\%} and {76\%} vs. {14\%} and {33\%}, respectively) than analytic dashboards. Differences between dashboard goals (\textit{motivation \& learning} vs. \textit{decision making}) are even more pronounced, for \typ{content blocks} ({93\%} vs. {49\%}), \typ{chart annotations} ({32\%} vs. {12\%}) and LV3-LV4 semantic levels ({58\%} and {75\%} vs. {8\%} and {29\%}). 

Many experts (5/13) report on a similar divide on the design of enterprise dashboards as  ``\textit{meant to be exploratory and analytical rather than narrative-driven}'' ($P9$), and they acknowledge actively avoiding ``\textit{prescribed analytics}'' ($P4$) in these cases: ``\textit{the last thing I want to do is to feed conclusions to the user}'' ($P12$). That said, there is still a place for LV4 content in corporate dashboards, in the form of metric definitions and explanations ($P4$, $P7$, $P9$, $P12$), 
metadata (e.g., when last updated) ($P7$), data alerts ($P4$), and interaction guidance ($P1$, $P5$): ``\textit{You're supposed to guide them on how to use the dashboard but not guiding them to the insights}'' ($P5$).

\implications{Context of use is a key factor guiding the choice of text elements and semantic level coverage. Dashboard genres, dashboard goals, dashboard audience, and frequency of use are potentially useful signals to discriminate text use at scale.}

\subsubsection{Text Formatting for Analytical Support and Emphasis}
\label{sec:practices-formatting}

In our coding exercise, we found that text formatting and placement is frequently used to support analytical reasoning. 
From interviews, we found that text formatting was the aspect of text use that experts were most intentional about, and the majority (7/13) reported following some form of text formatting guideline. The practices we compiled are: 

\begin{itemize}[noitemsep]

\item \noindent Leveraging font sizes to convey \textit{hierarchy of text components} (e.g., \typ{titles} vs. \typ{chart captions}) and support reading order ($P3$,$P4$).

\item \noindent Text color used as \textit{color legends} ($P6$, $P7$, $P9$) (Figs.~\ref{fig:malaysia_dashboard} and \ref{fig:school_finance_dashboard}).

\item \noindent Boldface and text color for \textit{emphasis} (e.g., data values, data labels, takeaways) ($P7$, $P8$);  (Figs.~\ref{fig:social_distancing_dashboard} and \ref{fig:malaysia_dashboard}).

\item \noindent Italics and bold face to \textit{distinguish} \typ{interaction guidance} from \typ{content blocks} ($P13$) (Fig.~\ref{fig:school_finance_dashboard}).

\item \noindent \textit{Avoid small fonts}, for legibility and accessibility reasons (5/13): ``\textit{screen readers can read best anything that's 12 pt and up}'' ($P4$)

\item \noindent \textit{Consistent use} of fonts (4/13): ``\textit{I don't like mixing fonts}''~($P4$).

\end{itemize}


While seemingly obvious, these text practices require deliberate action from dashboard designers. There is also a general consensus that text formatting affordances in dashboard tools are minimal, and scriptable tasks like aligning text color to match visual marks require much manual intervention. Unsurprisingly, several experts reported the use of \textit{style guides} in their corporate practice ($P4$, $P7$, $P11$) to help inform and align formatting practices across larger design teams.
 
\implications{Text formatting plays a significant role in dashboard communication and accessibility, but realizing its potential presently requires designer knowledge and experience. Given reasonable consensus around formatting practices, rule-based approaches to facilitate enforcement of best practices might be feasible as a first step. More importantly, explicit text-data linking would enable useful formatting affordances.}


\subsubsection{Managing Content Overload with Details on Demand}
\label{sec:practices-details}


Regarding overall dashboard content, many experts expressed a desire to minimize ``\textit{big walls of text}'' ($P1$, $P11$, $P12$), following a general understanding that users ``\textit{tend to glance over}'' text content ($P1$, $P4$, $P8$, $P13$): ``\textit{my approach is to shorten everything that you can shorten}''~($P11$). These considerations are particularly relevant in the context of explanatory dashboards, where experts report being very deliberate in choosing what content to include and trying to ``\textit{keep it as relevant as possible}'' ($P7$) by focusing on the key takeaways ($P4$, $P6$, $P7$): ``\textit{I'm constantly questioning myself. Why am I not including this? what is the ulterior motive of why you're doing this?}'' ($P4$).

As for minimizing visible text content, the most cited strategy is \rev{providing} details on demand via \typ{tooltips} ($P1$, $P7$, $P9$, $P10$, $P13$). \rev{While dynamic views and tabs were occasionally used for this same purpose}, tooltips were prevalent and featured a broad range of implementations, including embedded visual reports (e.g., Fig.~\ref{fig:social_distancing_dashboard} 
\dbl{136}{https://public.tableau.com/app/profile/emily.de.padua/viz/AllByMyself/AllbyMyself}). 
Another notable tooltip adaptation is having content formatted as a templated sentence (seen in {9\%} of dashboards, e.g., Fig.~\ref{fig:sentence-tooltip}), aimed at simplifying the interpretation of data facts ($P8$, $P13$): ``\textit{if it's a less technical audience, then I'll try and turn it into a sentence}'' ($P13$). 



\implications{Choosing what content goes into dashboards is a nuanced task that leans on minimizing visible text. Content summarization, text templating tools, and more content nesting affordances would be helpful for authors. From a content generation perspective, ranking strategies should also be considered.}

\begin{figure}
    \centering
    \includegraphics[width=0.95\columnwidth,alt={Dashboard overlay with instruction callouts over dashboard elements. Examples: (a) "Overall performance: click to switch monitoring metrics". (b) "Update field refreshed daily at 12PM CT".}]{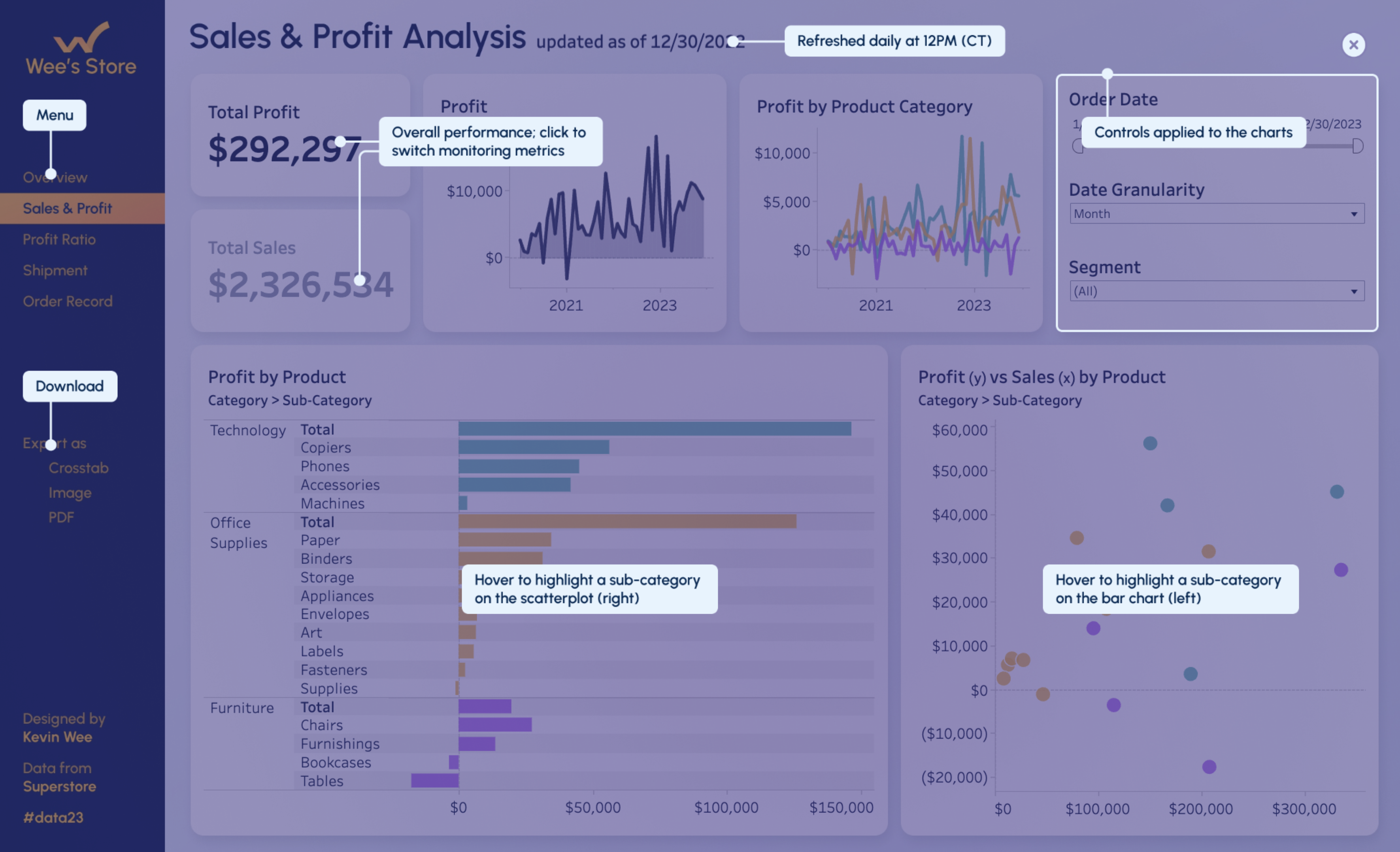}
    \caption{Dashboard depicted in Fig.~\ref{fig:transport_dashboard} \dbl{59}{https://public.tableau.com/app/profile/kevin.wee/viz/Data23DemoCreativeUsesofImagesinaTableauDashboard_16828225482900/Overview} with active instruction overlay.}
    \label{fig:instruction-overlay}
    \vspace{-1em}
\end{figure}

\subsubsection{Guiding Users (or Not)}
\label{sec:practices-guidance}


As discussed earlier, different dashboard genres entail different degrees of user guidance, from \textit{analytical support} that highlights salient points and articulates takeaways to \textit{usability support} via visualization interpretation instructions and interaction guidance. A lot of the text practices already covered pertain to user guidance, including the significant prevalence of \typ{interaction guidance} components ({58\%}), text formatting practices for guidance text, and simplifying analytical interpretation with sentence-style tooltips. 

Despite the differences in analytical guidance needs across dashboard genres, the presence of \typ{interaction guidance} in our corpus is fairly balanced across all non-\textit{static} genres (between {53\% and 65\%}). We found interaction guidance text to be most often displayed as co-located text blocks (either adjacent to or embedded in corresponding views) or as instructions in interactive widgets (e.g., filters). Other creative but less common implementations include: using icons in lieu of text 
\dbl{82}{https://public.tableau.com/views/MMW12CerealIndustry/MMW12CerealsIndustry};  \textit{instruction overlays} that, when active, reveals text instructions next to relevant dashboard components, e.g., 
\dbl{22}{https://community.fabric.microsoft.com/t5/Data-Stories-Gallery/COVID-19-NSW-Transport-Impact/td-p/1193474}, 
Fig.~\ref{fig:instruction-overlay} \dbl{59}{https://public.tableau.com/app/profile/kevin.wee/viz/Data23DemoCreativeUsesofImagesinaTableauDashboard_16828225482900/Overview}; and \textit{staged presentation}
\dbl{136}{https://public.tableau.com/app/profile/emily.de.padua/viz/AllByMyself/AllbyMyself}, where portions of the dashboard are shown or highlighted one at a time, e.g., 
\dbl{172}{https://community.fabric.microsoft.com/t5/Data-Stories-Gallery/Global-Stock-Market-World-s-Top-200-Companies/td-p/1482649}
\dbl{181}{https://community.fabric.microsoft.com/t5/Data-Stories-Gallery/Instagram-Year-in-Review-2017/td-p/337784}.

 While useful, guidance may not always be needed. Relevant use cases for guidance, according to experts, include audiences with lower visualization literacy ($P3$, $P13$) and bespoke or unusual interactions ($P6$, $P7$, $P13$): ``\textit{I like to assume most people know how general filters work, so I don't tend to explain those unless there is some sort of nuance that needs to be explained}'' ($P13$). One participant argues that explicit interaction guidance may not even be always beneficial: ``\textit{They're really useful, but they're only useful once, right?  (..) An area I continually struggle with is, do you make that explicit and make the experience for first-time users better, or do you not make that explicit and make it better for continuous users?}'' ($P9$). They also commented on cases where discovering hidden interactions makes for a ``\textit{delightful}'' experience: ``\textit{That's always fun to get the element of surprise. If it can be done well, you can spark joy}'' ($P9$).
 
\implications{ Determining appropriate levels of analytical and interaction guidance is a key design consideration that requires understanding the analytical needs and visualization literacy levels of the target audience. While guidance is not always necessary or desirable, there is value in providing better support for authors to create more and better guidance and encourage thinking about end users and their information needs.}

\subsubsection{Dynamic Text for Analytical Support and Breadcrumbs}
\label{sec:practices-dynamic}

One of the aspects considered in our coding exercise was the presence of dynamic text, i.e., understanding if and how text components respond to interaction or embody animations.
Dynamic text is present in {49\%} of all dashboards in our corpus and {62\%} of \textit{analytic} dashboards, underscoring its role in more exploratory-driven dashboards. 
Common versions of dynamic text include \typ{text-data summaries} that update with dashboard filters and interaction breadcrumbs for \typ{titles} and \typ{chart titles} that update to reflect newly selected data dimensions. These embodiments play an important awareness role in communicating what data slices are included or excluded, e.g., ``\textit{I'm putting on [filter indicators] on the dashboard so that they can see very clearly, `Oh, I have filters applied, I don't really mean to do that'} ''($P6$).   

Less common but interesting uses of dynamic text are on dynamic takeaways, i.e., pre-computed text findings that change depending on the filter selections (e.g., 
\dbl{53}{https://public.tableau.com/views/RWFDRetailSalesDashboard/Main} 
\dbl{149}{https://community.fabric.microsoft.com/t5/Data-Stories-Gallery/If-I-Were-a-Rich-Man-A-Bitcoin-What-If-Calculator/td-p/332301} 
\dbl{190}{http://saurabhr.com/wp-content/uploads/2020/06/TeamPaceAnalysis.html}); and a circular ``news ticker'' strips with data highlights (e.g.,
\dbl{22}{https://community.fabric.microsoft.com/t5/Data-Stories-Gallery/COVID-19-NSW-Transport-Impact/td-p/1193474}
\dbl{189}{https://arkadesignstudio.in/dataviz/}). 
On the latter, some experts commented on the value of animated features for user engagement ($P3$, $P7$, $P9$, $P10$): ``\textit{Maybe you hover on top of a tiny text and it's just goes bigger, little things that just keep you engaged}'' ($P7$); and some also mentioned practical uses of dynamic formatting and alerts in their corporate practice ($P4$, $P6$): ``\textit{if it's an overall data issue, it's at the very top and it gets color-coded by brand}'' ($P4$).
But despite the excitement and the meaningful use cases for dynamic text and dynamic formatting, adding such dynamic behaviors with mainstream dashboard tools is, again, reportedly challenging ($P2$, $P5$, $P6$, $P13$).

\implications{Dynamic text is a compelling resource to enhance comprehension and delight dashboard users but is difficult to implement with current tools. There are opportunities to provide more user-friendly ways to link dashboard text and data properties, and more flexible strategies for templated content generation.}


\subsubsection{Communication Elements Beyond Text}
\label{sec:practices-beyond}


Finally, we mention a few text-adjacent topics that have an impact on dashboard communication and that emerged in the coding and discussions. 
First, we noted a prevalence of icons for communication {(33\%)} either as space-efficient alternatives to text, as redundant visual elements, or as text decorators (e.g., Fig.~\ref{fig:school_finance_dashboard}). When well designed, icons and symbols enhance dashboard comprehension, but again, practices are not formalized. There is room to facilitate more principled icon usage via standard icon libraries and dedicated design recommendations.

Second, text plays an important role in dashboard \textit{accessibility}, as both an object (e.g., with formatting concerns) and enabler (e.g., via alt text) of accessibility. Experts agree that this is an important consideration, but also not a top-of-mind concern for most (7/13), and something that entails design trade-offs ($P9$) (e.g., limited font options). 

Third, an accessibility-adjacent topic that has received relatively little attention is language translation support. Our English corpus features two distinct uses for multilingual dashboards, including one where all text is duplicated to feature both English and Japanese 
\dbl{58}{https://public.tableau.com/views/SDGVizProject12/SDGS12}; and one featuring \rev{protest chants in Polish (with English }translations) for affect 
\dbl{78}{https://public.tableau.com/app/profile/zainab2225/viz/ProtestsagainstlimitingabortionrightsinPoland-MAPMM2021W8MakeoverMonday/Tojestwojna}. Advocating for the expansion of text visualization techniques and fostering multilingual accessibility is crucial for engaging a global audience and ensuring dashboard usability. 

\implications{While text is a major player in dashboard communication, many other communication aspects can affect or are affected by text. Keeping a broader outlook on the dashboard communication landscape may help inform additional areas of research.}





%% file: sections/04-fig-semantic-levels.tex
\begin{figure}[ht]
\centering
\includegraphics[width=.9\columnwidth, alt={Callout to text snippets in a dashboard featuring four levels of semantic detail. Level 1: 
This is a Gender \& Population graph. It shows the percentage of genders in New York City, and changes respectively to the chosen city. By hovering over the gender you will be able to see the total gender count. Level 2: #1 Population In America with 17M Population. Level 3: With 17M total population, 1 in every 38 people in the United States call New York City home. Level 4:  For students that wants to live in New York City affordably while exploring the city, it is difficult to find a place to stay. 
}]{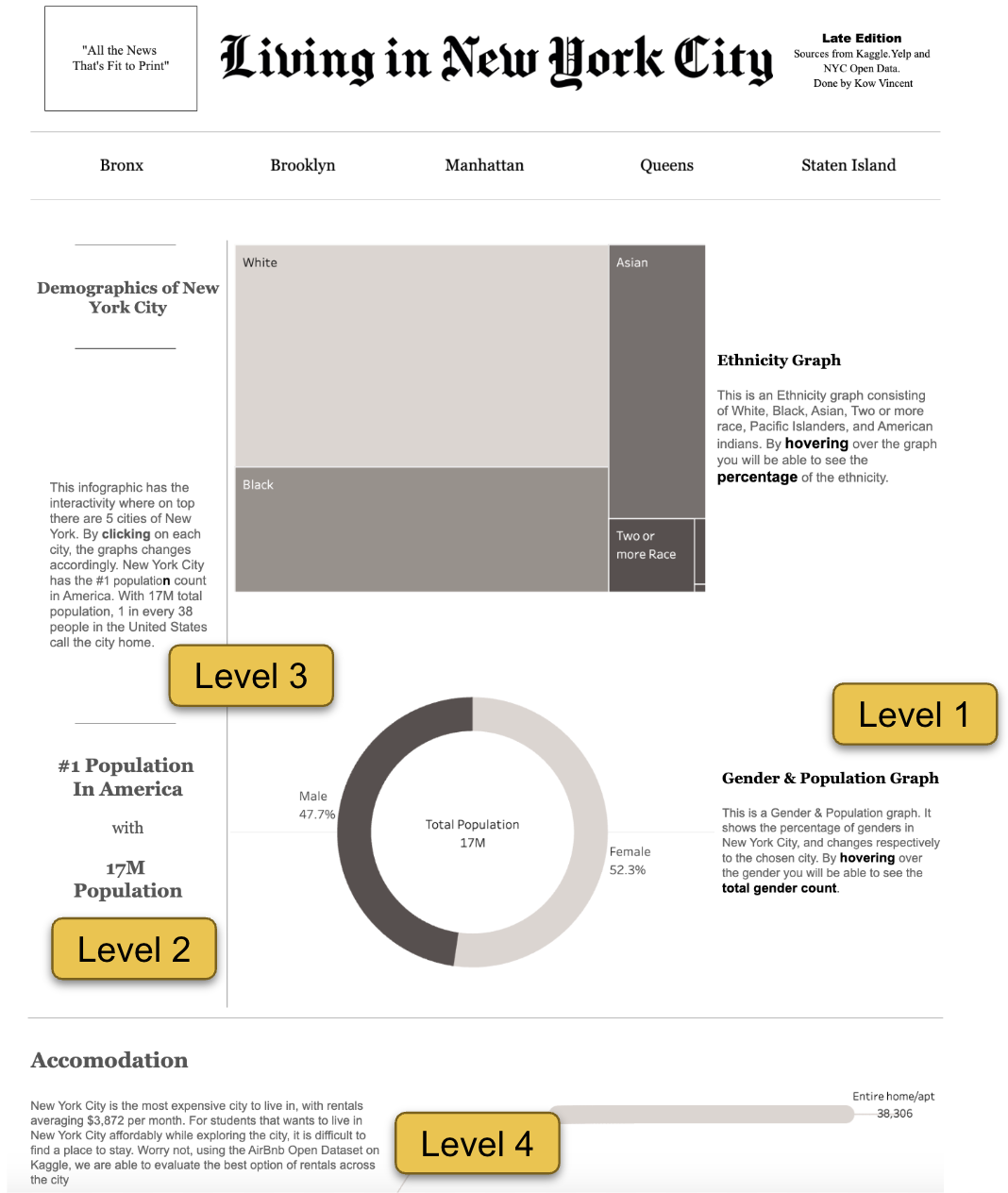}
\caption{A snippet of a dashboard displaying text components at varying levels of semantic detail \dbla{89}{https://public.tableau.com/views/Assignment1_16303281835750/Dashboard2}{Kow Vincent}{https://kowvincent.com/}. Level 1: basic chart descriptions and titles, indicating visible data dimensions like ``Gender \& Population Graph.'' Level 2: statistical concepts shown as BANs providing numerical summaries such as ``17M Total Population.'' Level 3: perceptual insights on population segments elucidate more complex data features. Level 4: contextual and domain-specific insights offer additional information regarding accommodation affordability in New York City.}
\label{fig:semanticlevels}
\vspace{-1.5em}
\end{figure}

%% file: sections/05-heuristics.tex
%

\begin{table*}[]
    \centering
    \begin{tabular}{ |c|c|p{0.75\linewidth}| }
    \hline
        \multirow{3}{*}{\textbf{\makecell{Analytical \\ Support}}}
         & \heurt{{\texttt{HA1}}} & \heurt{Text in the dashboard supports specific analytical questions or tasks.}
         \\
         & \heurt{{\texttt{HA2}}} & \heurt{The text in the dashboard should emphasize the most salient points of what the visuals in the dashboard convey.}
         \\
         & \heurt{{\texttt{HA3}}} & \heurt{The text in the dashboard supports readers in deriving clear takeaways from the dashboard.}
         \\
    \hline
        \multirow{3}{*}{\textbf{Semantics}}
        & \heurt{{\texttt{HS1}}} & \heurt{The text provides sufficient contextual information to describe what the dashboard is about.}
        \\
        & \heurt{{\texttt{HS2}}} & \heurt{Text content within the dashboard is at the appropriate level of detail to convey the intended message.}
        \\
        & \heurt{{\texttt{HS3}}} & \heurt{The text in the dashboard plays a role in disclosing sources, disclaimers, and biases.}
        \\
    \hline
        \multirow{4}{*}{\textbf{Presentation}}
        & \heurt{\texttt{HP1}} & \heurt{The text in the dashboard is legible, easy to read, and useful. Different parts of the dashboard are well-described.}
        \\
        & \heurt{\texttt{HP2}} & \heurt{Text supports a clear reading order within the dashboard, and it is logical.}
        \\ 
        & \heurt{\texttt{HP3}} & \heurt{Text formatting and placement provide a clear and consistent visual style, mood, and guidance in understanding the analysis.}
        \\
        & \heurt{\texttt{HP4}} & \heurt{Icons and symbols are used with text to help communicate patterns in data.}
        \\
    \hline
        \multirow{2}{*}{\textbf{Interaction}}
        & \heurt{\texttt{HN1}} & \heurt{Text makes it clear to the user where they need to start interacting with the dashboard.}
        \\
        & \heurt{\texttt{HN2}} & \heurt{The text indicates a clear path and breadcrumbs for performed user actions within the dashboard.}
        \\
    \hline
    \end{tabular}
    \caption{Our set of 12 heuristics for the use of text in dashboards, based on the work of Setlur et al.~\cite{setlur2023heuristics}.}
    \label{tab:heuristics}
    \vspace{-1.2em}
\end{table*}

\section{Recommendations for Text Use}
\label{sec:heuristics}

Reflecting on the empirical nature of text practices, we argue it is useful to materialize such tacit knowledge in the form of explicit design guidance, not only for dashboard authors but also for dashboard tool designers. As described in \S\ref{sec:coding-methodology}, we turned to the work of Setlur et al.~\cite{setlur2023heuristics} and their heuristics for analytical conversation in dashboards. We reframed their 39 heuristics from a text-centered perspective to derive a set of 12 heuristics (Table~\ref{tab:heuristics}) specifically on the use of text in dashboards to support analytical conversation. We present our derived heuristics (\S\ref{sec:heuristics-description}), discuss findings from applying these heuristics to our dashboard corpus (\S\ref{sec:heuristics-application}), and share expert impressions (\S\ref{sec:heuristics-validation}).


\subsection{Heuristics for Use of Text in Dashboards}
\label{sec:heuristics-description}

 Our construction process started by assessing each of the 39 heuristics in Setlur et al.~\cite{setlur2023heuristics} (\heid{H1}-\heid{H39}) and transforming relevant ones to be centered around text use. Some heuristics made explicit mentions of the appropriate use of text (e.g., \heid{H6} ``\textit{the text in the dashboard is legible, easy to read, and useful}'', here mapping to \heid{HP1}) and are used almost (if not) verbatim. Others covered communication aspects where text plays an implied or supporting role, which we then redrafted to explicitly mention text as an actor, e.g., \heid{H15} ``\textit{Charts within the dashboard are at the appropriate level of detail to convey the intended message}'', here mapping to \heid{HS2}, ``\textit{Text content within the dashboard are at the appropriate level of detail to convey the intended message}''. Heuristics with similar meaning after these redraftings were merged (e.g., \heid{H9} thru \heid{H12}, now \heid{HP3}). \rev{While we iteratively} fine-tuned descriptions and rubrics, the essence and scope of the heuristics remained.

Instead of the five conversation states used to organize the heuristics in Setlur et al.~\cite{setlur2023heuristics}, we grouped ours into four guidance categories tailored to more directly address properties and roles of text use in dashboards: \textbf{Analytical support}, referring to how text aids users in understanding the data and analytical questions they can derive from the dashboard; \textbf{Semantics}, referring to the role of context and granularity of information to support analytical tasks; \textbf{Presentation}, referring to formatting and layout considerations to support navigation; and \textbf{Interaction}, highlighting the role of interaction guidance and awareness.

\subsection{Applying the Heuristics}
\label{sec:heuristics-application}

We wanted to understand how heuristics may be used in practice to assess dashboards in supporting analytical conversation and what features of text use tend to contribute to each heuristic. As mentioned in \S\ref{sec:coding-methodology}, \rev{both} authors independently assessed dashboards in the corpus using the heuristics via a rubric that spans four assessment levels (\textit{strong application}, \textit{weak application}, \textit{weak violation} and \textit{strong violation}), 
and engaged in in-depth discussions to reach consensus on what aspects of dashboard text contributed to the assessment of each heuristic. We made our best efforts to account for differences in topic, genre, and dashboard goals in each case.
This process allowed us to shift from a creator's perspective to a user's perspective and to reflect on how text components in dashboards may best serve their intended audience. In the following segments, we share our process findings from applying the heuristics to our dashboard corpus and how our functional (\S\ref{sec:text-components}) and semantic (\S\ref{sec:semantic-levels}) categorizations may be leveraged to guide a heuristics-based dashboard assessment. For those seeking to perform similar heuristic assessments, we \rev{encourage the use of} our assessment rubric in our code book (Supplemental Materials).


\pheading{Analytical support.}
We found that having a combination of complementary text components that work well together is key for good analytical support. Header-level text blocks like \typ{title}, \typ{subheading}, \typ{section headers}, and \typ{chart titles} lend structure to connect data-level chart communication goals \heid{(HA1)} to higher-level takeaways \heid{(HA3)}. Some components make direct and obvious contributions to heuristics, such as \typ{chart annotations}, \typ{text-data summaries}, and \typ{content blocks} to salient points \heid{(HA2)}, and subheading to takeaways \heid{(HA3)}, but as discussed earlier in \S\ref{sec:practices-genres}, prescriptive analytics are not always needed or desired, \rev{nor} more content equate better communication. Several dashboards in our corpus showcased good application of these heuristics even with minimal use of text 
\dbl{53}{https://public.tableau.com/views/RWFDRetailSalesDashboard/Main} 
\dbl{98}{https://public.tableau.com/views/Health_Rate_Dashboard_16120785130490/Dashboard1} 
\dbl{111}{https://public.tableau.com/views/TablesTest2/Dashboard}.



\pheading{Semantics.}
Assessing sufficient context \heid{(HS1)} is a subjective measure that heavily depends on the use case and intended audience~\cite{hullman2011}. We assessed this metric based on the \rev{extent we} were able to interpret and analyze dashboard data given available information (e.g., in \typ{content blocks}, \typ{tooltips}, and \typ{chart annotations}). Dashboards for expert users or general knowledge topics (like sports or the pandemic) might be sufficiently grounded without explicit context (LV4), but we found that having explicit context was generally helpful. 
On the other hand, the assessment of semantic levels of detail \heid{(HS2)} is more objective and can be characterized in terms of the coverage of the LV1-LV4 content spectrum to support analytical reasoning. A strong application here means that there are clear reasoning paths from LV1 and LV2 findings to overarching conclusions (LV3 and LV4). All text components collectively play a role in supporting this aspect.
Finally, sources, disclaimers, and biases \heid{(HS3)} are present in \typ{metadata} blocks; in our assessment, applications featured easily retrievable sources {(45\%)}, and strong applications featured notes on data collection and processing~{(19\%)}.

\pheading{Presentation.}
Presentation factors \rev{like formatting} and placement play both functional and aesthetic roles in dashboard communication\rev{, and are addressed by \heid{HP2} and \heid{HP3}}. Good applications for \heid{HP2} refer to the extent \rev{that} dashboard text supports users \rev{with navigation}: it is about conveying a strong visual sense of content hierarchy via font size and header placement for content navigation, as well as strategic placement of text blocks close to relevant visual components. \heid{HP3}, on the other hand, focuses on appropriate and consistent use of text formatting more holistically, including fonts, color, mood, and branding. \heid{HP1} pertained specifically to the quantity and quality of text at appropriate levels to fulfill dashboard goals; dashboards we assessed lower in this criteria featured little to no content beyond \typ{chart titles}. Finally, \heid{HP4} is about ensuring that icons (when used) have a synergistic role with text, clearly communicating their intent and \rev{matching dashboard} aesthetics.


\pheading{Interaction.}
The final two heuristics pertain to the role of text in scaffolding interaction. The first is about aiding discovery \heid{(HN1)}, which we assessed as the presence of explicit text instructions (i.e., \typ{interaction guidance} blocks) to point new users to key interactive features of the dashboard; strong violations \rev{happened when} meaningful interactive features were missed due to lack of explicit guidance. The second refers to text communicating interaction outcomes \heid{(HN2)}, which was most frequently embodied as dynamic labels updating to match filters, e.g., a \typ{chart title} updating alongside an updated chart. Strong violations \rev{here map} to cases where no traces of user action are provided in text or visuals, \rev{e.g., filtered views misrepresented as whole dataset views.}


\subsection{Expert Feedback on Heuristics}
\label{sec:heuristics-validation}

Expert feedback on heuristics was largely positive. The empirical nature of text practices is again acknowledged ($P4$, $P6$, $P12$): e.g., ``\textit{I teach data visualization frequently (...) [and] we don't really talk about text that much, or how to use text appropriately}'' ($P6$); and that having these practices formalized as heuristics is useful (10/13):  ``\textit{For those who are not familiar with the norms or may come from an experience where the norms are different, [this formalization] is a unifier}'' ($P4$). Cited benefits and uses entailed: fostering intentional and ``\textit{thoughtful usage of text}'' ($P5$, $P6$, $P13$); guidance for novice designers ($P6$, $P8$) and sanity checks for experienced designers ($P5$, $P9$);  and providing a base for more bespoke guidance in applied scenarios ($P7$). Participants also stated that our set of heuristics felt complete (8/13) and representative of their practices (5/13), and they also appreciated the structure with four guidance categories ($P3$, $P4$, $P7$). 

While the heuristics were found to provide ``\textit{a good template}'' ($P7$) for recommended text practices, many participants argued more specificity is needed to support practical hands-on support ($P3$, $P4$, $P8$, $P9$), e.g., ``\textit{If it's just guidance to point people in the right direction, great. But if you want people to have a standard and follow, I need a little more functional direction and just something to grasp and hold on to}''~($P4$). A participant also remarked that the heuristics on their own will not ensure dashboard quality, e.g., ``\textit{you tick the boxes altogether and then maybe you might get an ideal dashboard, but sometimes you might not}'' ($P12$). Topic suggestions for more specific guidance entailed best practices in text formatting~($P4$, $P5$, $P7$), guidance for annotations ($P3$), and an example catalog to showcase both good and bad applications of individual heuristics ($P3$, $P5$, $P11$, $P13$). More generally, some also underscored the importance of ``\textit{context and perspective of the user}'' ($P10$, $P12$), and better communicate the extent to which each heuristic matters on different use cases ($P10$).

%% file: sections/06-future.tex
\section{Opportunities for Text use in Dashboards}
\label{sec:opportunities}
We identified several areas for future research that build on our dashboard findings, discussions with experts, and application of heuristics. 

\pheading{Dashboard linters.} 
Based on the strongly empirical nature of dashboard text practices (\S\ref{sec:practices-experience}) and informal dashboard text practices (\S\ref{sec:practices-formatting}), novice dashboard designers would greatly benefit from direct design guidance. Akin to the goals of visualization linters~\cite{mcnutt2018linting,chenvizlinter}, dashboard linters could provide real-time feedback on text use during authoring, ensuring adherence to best practices within the scope of our 12 heuristics (\S\ref{sec:heuristics}), e.g., typography, readability, and information hierarchy.
These tools could also potentially offer suggestions for improving text clarity\rev{, as well as accessibility~\cite{srinivasan2023azimuth}}.

\pheading{Example-based style transfer and recommendations.}
In view of the prevalent use of style guides in large design teams (\S\ref{sec:practices-formatting}), we also see value in more easily transferring text design features from one dashboard to another, similar to the notion of style transfer in image rendering~\cite{jing2019neural}. By analyzing exemplary dashboards within various domains and genres, algorithms can also provide smart defaults and recommendations for effective text formatting styles (e.g., font sizes, colors, hierarchy) that can help lower the barriers for text design that can effectively support navigation and reading order (\S\ref{sec:practices-experience}, \S\ref{sec:practices-semantic}).





\pheading{Automatic explanations.} 
Reflecting on user guidance needs for exploratory and explanatory dashboards alike (\S\ref{sec:practices-genres}, \S\ref{sec:practices-guidance}), we see an important opportunity for the development of automated and semi-automated content generation tools for context-specific explanations on underlying data. In the context of corporate dashboards, for example, explanations on visible metrics are widely used but onerous to create and keep consistent with changes in the data processing pipeline. Being able to generate accessible, domain-specific explanations that can be vetted by dashboard authors would significantly ease the burden on the author to produce comprehensive explanations.

\pheading{Provisions for text+charts links in dashboard design.} 
\rev{Following research on text-chart links for visual storytelling~\cite{sultanum2021leveraging, metoyer2018coupling, latif2021deeper, zheng2022evaluating, zhi2019linking, kwon2014visjockey},} facilitating the linking of dashboard text and charts would allow for exciting opportunities to expand the capabilities of dynamic text for analytical communication and storytelling (\S\ref{sec:practices-formatting}, \S\ref{sec:practices-dynamic}). 
Automating the linking of text and charts in this context would involve developing algorithms that not only detect and establish connections between textual descriptions and corresponding data points for \rev{guidance~\cite{latif2021kori,metoyer2018coupling, pinheiro2022charttext}}, but also suggest contextually relevant \rev{emphasis~\cite{strobelt2015guidelines}}, such as bolding key takeaways, coloring data references, or italicizing instructions (\S\ref{sec:practices-formatting}, \S\ref{sec:practices-guidance}).
Future research could focus on creating intelligent storytelling tools that adapt text formatting based on the data narrative and integrate these capabilities into existing dashboard authoring platforms.


\pheading{Guidance overlays and personalization.} 
Incorporating instruction overlays into dashboards can guide users through complex datasets or analysis features, enhancing learnability (\S\ref{sec:practices-guidance}). Drawing from just-in-time learning and in-situ instructions~\cite{STOIBER202268}, research could explore the inclusion of overlays that adapt to user preferences and interaction patterns, offering better scaffolding for dashboard users. By analyzing user behavior, along with analytical and accessibility needs (\S\ref{sec:practices-genres}, \S\ref{sec:practices-beyond}), the dashboard could then take user assistance a step further by tailoring personalized walkthrough versions for specific audiences, featuring data insights relevant to the user's specific needs or roles.

\pheading{Reactive documents meet dashboards.} \rev{Reflecting on next steps for dynamic text }(\S\ref{sec:practices-dynamic}) and the benefits of prose text for analytical understanding~(\S\ref{sec:practices-details}), we are inspired by past work on explorable documents ~\cite{dragicevic2019increasing, victor2011scientific}. Applying these concepts to interactive \rev{dashboards presents a fertile ground for future research where }interactive narratives and user-driven data exploration coalesce. For example, revealing textual explanations based on user \rev{interaction; breadcrumbs that provide dynamic summaries of the analytical journey; and contextual information surfacing on demand}. Future work could also explore how text can play an active role in the storytelling experience through interactive \rev{text widgets that allow users to manipulate data directly from the narrative. }



    

%% file: sections/07-limitations.tex
\section{Limitations}

While we made deliberate efforts to increase diversity and representation in our dashboard sample, our research primarily analyzes publicly available \rev{dashboards, a limitation} given the vast universe of private corporate dashboards that remain beyond our purview. Business \rev{dashboards in} public platforms (e.g., Tableau Public) are \rev{often} portfolio versions depicting mock data, whereas real business dashboards are embedded in intricate data ecosystems\rev{, feature complex visualizations, and} tend to be tailored for specific organizational contexts and decision-making processes, potentially incorporating proprietary design elements and interaction paradigms. 
This exclusion could mean that our analysis misses crucial insights into dashboard design and use in professional environments and might raise concerns regarding the generalizability of our findings. While our interview experts helped us uncover various text-related concerns relevant to their professional practice, we argue that collaborations with professional and media organizations could help identify novel applications of text in domain-specific use cases, thus broadening the applicability of this~work.

We also acknowledge the inherent subjectivity that accompanies our evaluative framework of the heuristics. 
Arriving at a unanimous consensus on the application or violation of heuristics was a laborious and nuanced exercise, but also an informative one, and we stand by the value of these heuristics as tools for design guidance.
That said, we agree that further assessment with wider usage would be valuable, e.g., evaluating the heuristics with novices as per original work~\cite{setlur2023heuristics}.

Finally, while the heuristics proposed in our work provide a structured approach to evaluating and designing text for dashboards, they may not offer the specificity required for direct application in all authoring scenarios. This limitation arises partly from the diverse contexts in which dashboards operate, making it challenging to formulate universally applicable rules without sacrificing relevance to specific analytical use cases and target audiences. However, the broad nature of our guidelines also presents an opportunity for designers to interpret these guidelines in ways that best suit their unique objectives and user needs.

    

%% file: sections/08-conclusion.tex
\section{Conclusion}
Our work highlights the fundamental role of text in the context of dashboard design and communication, wherein textual elements do much more than accompany visualizations; they enrich and clarify, providing essential context and direction. We hope this paper provides useful insights and future directions for \rev{not only dashboard authoring tools to better leverage the utility of text with visualizations, but also to guide effective analytical conversation and text generation in dashboard-adjacent text-data spaces, such as data stories~\cite{shin2022roslingifier, shi2020calliope}, data videos~\cite{xu2023end}, computational notebooks~\cite{li2023notable}, visualization thumbnails~\cite{kim2023towards}, and chart captioning~\cite{liu2020autocaption, srinivasan2023azimuth}.}
This synergy between text and visual\rev{s} reaffirms the adage, ``A picture is worth a thousand words, but a word can paint a thousand pictures'', underscoring the dual power of text to distill complex data into information understanding.